\documentclass[twocolumn,aps]{revtex4}

\begin{document}


\title{Reply to ``Comment on `Properties of the massive Thirring
  model\\ from the $\mathbf{XYZ}$ spin chain' "}

\author{M. Kolanovi\'{c}}\email{mk679@nyu.edu}
\affiliation{Center for Cosmology and Particle Physics,
  Department of Physics\\ New York University, New York, NY 10003}
\author{S. Pallua}\email{pallua@phy.hr}
\author{P. Prester}\email{pprester@phy.hr}
\affiliation{Department of Theoretical Physics, PMF, University of
  Zagreb\\ Bijeni\v{c}ka c. 32, p.p. 331, 10000 Zagreb, Croatia}

\date{February 12, 2003}

\begin{abstract}
We elaborate in more details why lattice calculation in
[Kolanovic et al, Phys.\ Rev.\ D {\bf 62}, 025021 (2000)] was done
correctly and argue that incresing the number of sites is not expected
to change our conclusions on the mass spectrum.
\end{abstract}


\maketitle

In the comment on our paper \cite{PP00} Fujita, Kobayashi and
Takahashi claim that our results for the mass spectrum of the massive
Thirring model (MTM) are not reliable \cite{FuKoTa03}. In particular,
they claim that if one uses spin chain regularisation it is
\emph{necessary} to diagonalize the spin chain Hamiltonian with a
number of sites $N$ larger than 1000. We now explain why we disagree
with this criticism.

The criticism of Fujita et al. is based on the following argument. If one
makes the standard spin chain regularisation of MTM (which is $XYZ$ spin
1/2 chain \cite{Lut,Lusch}) to obtain some reasonable results on
continuum extrapolation, one has to satisfy the condition
\begin{equation} \label{fcont}
\frac{2\pi}{N} \ll a m_0 \ll 2\pi.
\end{equation}
Here $N$ is the number of sites, $a$ is the lattice spacing, and $m_0$ is
a bare mass parameter. Using (\ref{fcont}) Fujita et al. claim that if
one wants to obtain any reliable information on the bound state of the
MTM one has to take $N>1000$, which is much larger than ours $N\le 16$.
Moreover, they claim that for values of the parameters used in
\cite{PP00} the left inequality in (\ref{fcont}) is even completely
violated, i.e., $2\pi/N>am_0$. In addition they claim that the mass
gap in our calculation is approximately equal to the ``resolution''
$2\pi/L$. From all this, Fujita et al. conclude that our results
\cite{PP00} cannot be very reliable for the bound state spectrum of
the massive Thirring model.

Let us now review standard lattice philosophy. Continuum regime in
lattice calculations is obtained when the correlation length $\xi$ is much
larger than the lattice spacing $a$ and at the same time much smaller than
spatial extension $L=Na$, i.e.,
\begin{equation} \label{corr}
a \ll \xi \ll Na \;.
\end{equation}
As $\xi=1/M$, where $M$ is the mass gap (mass of the lightest
particle), (\ref{corr}) can be equivalently written as
\begin{equation} \label{cont}
\frac{1}{N} \ll Ma \ll 1
\end{equation}
Now, hardware limitations make restrictictions on $N$. For example, in
lattice (quenched) QCD maximum lattices which are presently calculable
have $N\le64$ (see e.g., \cite{aoki}). More importantly here, for
exact diagonalisation in two dimensions $N<30$. It follows that
``$\ll$'' in (\ref{cont}) at best means 5--8 times smaller. So in
practical calculations one effectively imposes the condition
\begin{equation} \label{rcont}
\frac{1}{N} < Ma < 1
\end{equation}
and from the quality of scaling and the accuracy of the continuum
extrapolation (usually using different methods as a check) one decides
weather the extrapolated results are a good approximation of the
continuum theory. Of course, one cannot exclude the possibility that
for values of scaling parameter larger than accesed there is a
complete change of scaling behavior so that the obtained extrapolated
results are wrong\footnote{For nice discusion see Sec. 9.5 in Ref.
\cite{ChrHen}.}.

Let us now apply the above textbook analysis to our lattice
calculation ($XYZ$ spin chain regularisation of MTM) \cite{PP00}. For
the sake of clarity we restrict ourself to a interval of coupling
constant where elementary fermion is the particle with the lowest
mass. First, it is easy to see that (\ref{cont}) is \emph{not} equal
to (\ref{fcont}) used by Fujita et al.
It was shown in \cite{Lusch} that (using notation from \cite{PP00})
\begin{displaymath}
m_0a=\frac{8\gamma}{\pi}\sin\gamma
 \left(\frac{Ma}{4}\right)^{2\gamma/\pi}\;.
\end{displaymath}
In particular, let us analyze the left inequality in (\ref{rcont}). In our
extrapolation we had $Ma>0.2$ which is for $N=16$ reasonably larger
than $1/N=0.06$. It follows that in our analysis continuum condition
is fairly satisfied when \emph{proper} condition is used, contrary
to claim of Fujita et al. 

As we mentioned above, there is always the possibility that for larger
$N$ something dramatic happens with the scaling law and that our
extrapolations made with $N\le16$ are incorrect. There are several
reasons why we believe that this should not be expected in the case of the
$XYZ$ chain:
\begin{itemize}
\item
global properties of the energy spectrum are as expected for QFT with
calculated mass spectrum. In particular, there are states with
energies corresponding to two-fermion and two-(first)breather states
\item
the same technique was previously succesfully applied for similar
perturbed CFT's, e.g. ordinary and tricritical Ising model in a
magnetic field \cite{Henkel}. It was shown there that passing from
$N\le14$ to $N\le21$ only slightly improved precision and conclusions
about the spectra remained the same. In fact it was shown that very
accurate results can be obtained already for lattices with $N\le24$
\cite{CasHas}.
\item
our results agree with the DHN mass formula \cite{dhn} obtained by a
number of different methods. It is very hard to imagine how bad
extrapolation can agree with exact analytic result. We should also
mention that a similar analysis for lattice regularization of the
sine-Gordon model (periodic $XXZ$ chain in transverse magnetic field)
also gave results consistent with the DHN formula \cite{PP99}.
\item
Beside the mass ratios in $L\to\infty$ limit we obtained anomalous
dimensions of corresponding states, in particular we had for the first
time calculated dimension of the second breather. Our result was
subsequently confirmed by analytical calculation \cite{FRT99}.

\end{itemize}

We conclude that our numerical analysis made in \cite{PP00} was done
correctly and that all results should be trusted, in a sense that
enlarging lattice would just increase numerical precision and leave
all conclusions unchanged.

Finally, we would like to comment on the Bethe ansatz solution of the
MTM. Spectrum of the MTM can be found by solving the Bethe ansatz 
equations in the continuum approximation \cite{bt}. The spectrum found
there agrees with the results of \cite{dhn}. Recently, it was argued
\cite{fu} that there exist no complex solution to the Bethe equations
and that there is only one bound state.
In \cite{bt} precisely the complex solutions (so called Bethe strings)
are describing bound states of MTM. We performed numerical analysis
of Bethe equations for MTM and tried to reproduce the result of
\cite{fu}. Our results indicate that the powers $\alpha$ that
determine the scaling behavior of energies of the few lowest states on
the density $\rho$ differ and even vary with the $\rho$ (for
definitions see section 4 in \cite{fu}). Therefore we were unable to
extract conclusive results by letting $\rho\longrightarrow \infty$. 
An indication that the numerical iterations might easily miss the
complex solutions (that nevertheless exist) is coming from our study
of the Bethe ansatz equations for different spin chains
\cite{strings}. There we studied how the complex (string) solution in
two and three particle sectors emerge and disappear when one changes
the parameters of the model. We found that iterative numerical methods
fail to converge on string solutions, although they exist and can
be found analytically. 

\begin{acknowledgments}

MK would like to thank to J.H. Lowenstein for useful and
stimulating discussion.

\end{acknowledgments}


\begin{thebibliography}{99}

\bibitem{PP00}
  M. Kolanovi\'{c}, S. Pallua and P. Prester, Phys.\ Rev.\ D {\bf 62},
    025021 (2000).
\bibitem{FuKoTa03}
  T. Fujita, T. Kobayashi and H. Takahashi, hep-th/0306173.
\bibitem{Lut}
  A. Luther, Phys.\ Rev.\ B {\bf 14}, 2153 (1976).
\bibitem{Lusch}
  M. L\"{u}scher, Nucl.\ Phys.\ {\bf B117}, 475 (1976).
\bibitem{aoki}
  S. Aoki et al. (CP-PACS Collab.), hep-lat/0206009.
\bibitem{ChrHen}
  P. Christe and M. Henkel, \emph{Introduction to Conformal Invariance
    and Its Applications to Critical Phenomena} (Springer-Verlag,
    Heidelberg, 1993).
\bibitem{Henkel}
  M. Henkel and H. Saleur, J.\ Phys.\ A {\bf 22}, L513 (1989);
  M. Henkel, J.\ Phys.\ A {\bf 24}, L133 (1991);
  M. Henkel, J.\ Phys.\ A {\bf 23}, 4369 (1990).  
\bibitem{CasHas}
  M. Caselle, M. Hasenbusch, A. Pelissetto and E. Vicari,
    J.\ Phys.\ A {\bf 33}, 8171 (2000);
  M. Caselle and M. Hasenbusch, Nucl.\ Phys.\ {\bf B579}, 667 (2000).
\bibitem{dhn}
  R. F. Dashen, B. Hasslacher and A. Neveu, Phys.\ Rev.\ D {\bf 11},
    3424 (1975).
\bibitem{PP99}
  S. Pallua and P. Prester, Phys.\ Rev.\ D {\bf 59}, 125006 (1999);
  S. Pallua and P. Prester, Fiz.\ {\bf B10}, 175 (2001).
\bibitem{FRT99}
  G. Feverati, F. Ravanini, and G. Takacs, Nucl.\ Phys.\ {\bf B570},
    615 (2000).
\bibitem{bt}
  H. Bergknoff and H. B. Thacker, Phys.\ Rev.\ D {\bf 19}, 3666 
    (1979);
  H. Bergknoff and H. B. Thacker, Phys.\ Rev.\ Lett.\ {\bf 42}, 135
    (1979).
\bibitem{fu}
  T. Fujita, Y. Sekiguchi and K. Yamamoto, Annals Phys.\  {\bf 255},
    204 (1997).
\bibitem{strings}
  A. Ilakovac, M. Kolanovi\'{c}, S. Pallua and P. Prester,
    Phys.\ Rev.\ B {\bf 60}, 7271 (1999);
  A. Ilakovac, M. Kolanovi\'{c}, S. Pallua and P. Prester,
    Fiz.\ {\bf B8}, 453 (1999).

\end{thebibliography}
\end{document}